# Multi-mode fiber enabled multi-wavelength optical trapping and dynamic manipulation

Yikun Shen[1], Chuangye Zhang[1] , Yuquan Zhang[1], Jiahui Pan[1], Siwei Chen[1], Hongyang Xu[1], Yixuan Chen[1], Qi Jin[1], Xiaocong Yuan[1,*], and Changjun Min[1,*]

[1]*Nanophotonics Research Centre, Institute of Microscale Optoelectronics & State Key Laboratory of Radio Frequency Heterogeneous Integration*

**Address correspondence to: cjmin@szu.edu.cn and xcyuan@szu.edu.cn*

**Abstract:** Optical fiber tweezers offer distinct advantages for long-distance manipulation, compact integration, and minimally invasive operation in biological environments. However, most optical fiber tweezers rely on single-mode fibers (SMFs), which are constrained by limited optical mode diversity and reduced control flexibility. Although multi-mode fibers (MMFs) support a wider spectrum of propagation modes, their inherent mixed guided modes with low coherence become a long-standing limitation for the design of focused trapping configurations. To address these limitations, we propose and experimentally validate a fully MMF-based optical tweezer system integrated with a micro-lens structure fabricated on the fiber facet, enabling stable optical trapping across multiple wavelengths and dynamic manipulation of trapped cells. Employing 532 nm continuous-wave and 800 nm femtosecond lasers, we demonstrate that both light sources can generate tightly focused optical spots through the micro-lens with a high numerical aperture (NA > 0.7), achieving robust trapping and axial dynamic manipulation of cells. Compared with conventional SMF-based tweezers, this approach leverages the broadband and multi-mode properties of MMFs, allows for wavelength-flexible and dynamically adjustable trapping of cells, and paves the way for lab-on-fiber biophotonic platforms with potential applications such as interventional manipulation, cell sorting, and cellular fluorescence analysis.

**Keywords:** optical fiber tweezers; multi-mode fiber; optical trapping; micro-lens; cell manipulation

## 1. Introduction

Optical tweezers have emerged as a fundamental technology in modern biophotonics, enabling non-contact manipulation of cells, organelles, and micro- to nano-scale particles, and become essential tools for conducting precise investigations in the life sciences [1–4]. Among various configurations, fiber-based optical tweezers have attracted considerable attention due to their compact design, ease of integration, and favorable biocompatibility. In 1993, Constable et al. [5] first demonstrated particle trapping by counter-aligning two optical fibers to balance scattering forces. Subsequently, Taguchi and colleagues later advanced this technology by developing a single-fiber optical tweezer system based on tapered fibers, successfully trapping polystyrene microspheres and live yeast cells in liquid environments [6]. More recently, efforts to develop compact and integrated optical tweezer platforms have yielded notable progress. Yuan's group investigated the use of single-mode fibers (SMFs) with fusion tapering to achieve three-dimensional trapping [7] and controlled transport of yeast cells [8]. Concurrently, Xin's group developed precise manipulation strategies using tapered fiber probes, enabling stable trapping, guided transport, and on-demand spatial arrangement of both synthetic particles and biological specimens [9]. Despite these advancements, conventional fiber-based optical tweezers remain fundamentally constrained by their dependence on SMFs. The small core diameter of SMFs supports only a single propagation mode, resulting in a fixed focal spot near the fiber tip with limited numerical aperture (NA) and low trap stiffness [10]. This restricts dynamic control over trapped objects and reduce system versatility. Furthermore, SMF-based tweezers are generally optimized for single-wavelength operation, restricting the capacity to tailor optical forces or minimize photodamage across diverse biological samples, thereby limiting their broader applicability in life science researches [11–13].

In contrast to SMFs, multi-mode fibers (MMFs) feature larger core diameters and support multiple propagation modes, thus offering new potential for achieving high-NA trapping and dynamic manipulation of samples in fiber-based optical tweezers. However, the presence of mixed guided modes within MMFs leads to low coherence and random phases in the output beam, therefore it is a long-standing and difficult problem to design tightly focused trapping configurations with MMFs. Recent studies have investigated hybrid configurations integrating MMFs with SMFs [14,15], wherein the SMF typically serves as the input channel and a spliced MMF segment is employed for mode expansion and spatial beam shaping. By incorporating focusing structures on the fiber facet, such systems can generate high-NA focal fields for trapping and axial movement of particles. However, these hybrid approaches fundamentally depend on SMFs as the primary transmission medium, relegating the MMF to a short mode-expansion segment, thus do not fully leverage the inherent advantages of MMFs, particularly their mode diversity and broadband operation capabilities. Consequently, the development of fully MMF-based optical tweezers remains a considerable challenge.

In this work, we propose and experimentally demonstrate a fully MMF-based optical tweezer system that enables high-NA focusing, three-dimensional trapping, and dynamic manipulation of biological cells under multi-wavelength operations. By designing and fabricating a micro-lens structure on the output facet of the MMF, high-NA optical foci are generated for different wavelengths. Employing representative 532 nm continuous-wave and 800 nm femtosecond pulsed laser sources, we systematically investigate the trapping performance for each wavelength at and demonstrate stable three-dimensional trapping of yeast cells. Moreover, by modulating the lateral coupling position of the incident light at the fiber input facet, controllable axial displacement of the trapping focus is achieved, enabling dynamic manipulation of the trapped cells. Compared to conventional SMF-based tweezers, the proposed MMF tweezers system offers several intrinsic advantages. First, the large core diameter and multiple propagation modes of the MMF effectively increase the focusing NA (NA > 0.7), significantly enhancing the optical gradient forces for trapping. Second, the MMF supports broadband wavelength transmission, enabling stable multi-wavelength trapping within a single platform. Additionally, precise control over the input coupling conditions provides further degrees of freedom for real-time manipulation of trapped samples. These features highlight the system's optical field tunability and spectral versatility, thereby opening new prospects for applications such as high-throughput biological manipulation [16], single-cell motility and size sensing [17,18], in-flow trapping in microfluidic environments, and fiber-based endoscopic imaging [19,20].

## 2. Operating Principle and System Design

### 2.1 Experimental Setup and Micro-Lens Design

The operational principle of the MMF optical tweezers system is schematically illustrated in Fig. 1(a). Initially, an incident beam is focused to couple into the input facet of the MMF, thereby inducing multi-mode interference within the fiber core. This interference produces an annular output field at the distal end of the fiber, with a tunable radius. After being refocused by the integrated micro-lens structure on the fiber end-face, the emitted light forms a high-NA focal spot, generating a stable three-dimensional optical trap capable of confining cells and other microscopic specimens.

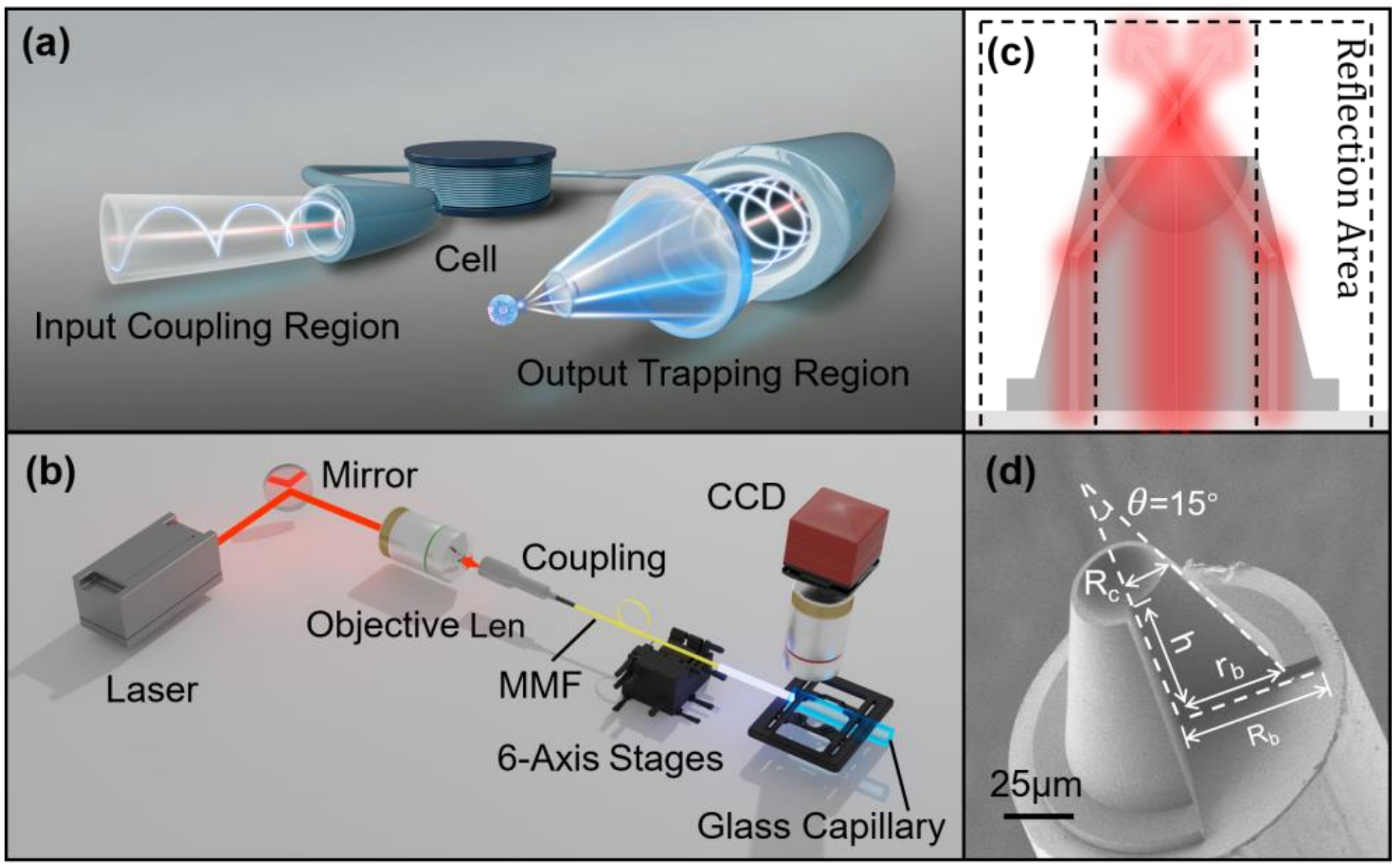

Fig. 1: The proposed MMF optical tweezers system and the micro-lens design. (a) Schematic of the operating principle. Incident light sources are coupled into the MMF, where multi-mode interference within the core generates annular output beams. Upon refocusing by the end-face micro-lens, the focus beyond the MMF tip forms a three-dimensional optical potential well, enabling stable cell trapping. (b) Schematic of the experimental setup, which consists of the MMF, six-axis positioning stages, LED illumination, and a CCD-based imaging system. (c) Design of the micro-lens structure on the fiber output facet. The laterally sloping regions reflect and redirect the emitted light toward the optical axis, increasing the convergence angle and effective NA, thus enhancing the trapping potential. (d) Scanning electron microscope (SEM) image of the fabricated micro-lens, revealing an axisymmetric truncated-cone geometry with a hemispherical concave cavity at its apex.

As shown in Fig. 1(b), the proposed MMF optical tweezer system comprises two primary modules: a beam-coupling unit and a cell manipulation platform. The beam-coupling unit is designed to efficiently launch the incident laser beam into the MMF. To evaluate the multi-wavelength trapping capability and broadband compatibility of the MMF, two different laser sources were employed: a 532 nm continuous-wave laser (MSL-FN-532-S, 400 mW, Changchun New Industries) and an 800 nm femtosecond laser (Coherent Legend Elite HE, 300 mW, 100 fs pulse duration, 80 MHz repetition rate). Due to the large number of guided modes supported over a broad spectral range, each wavelength inherently can excite a distinct superposition of modes inside the MMF. In the system, the laser beam is first directed through a neutral density filter to adjust its power, then aligned using reflective mirrors to ensure a stable, high-quality beam profile. It is subsequently focused by a 20× objective lens (Nikon, NA = 0.5) and coupled into a MMF (Shenzhen Optics-Forest Co. Ltd., cladding diameter 125 μm, core diameter 105 μm). A micro-lens fabricated on the fiber output facet concentrates the multi-mode interference field into a tightly confined, high-NA focal region. The resulting focal intensity distribution forms a stable three-dimensional optical potential well that reliably traps microscopic objects such as yeast cells. To minimize disturbances caused by liquid surface tension, the trapping platform incorporates a six-axis precision translation stage that controls the three-dimensional movement of the fiber tip within a capillary glass tube (cross section 0.7 × 0.7 $mm^2$, length 30 mm). Uniform white-light illumination is provided from below by an LED source. The trapping process at the fiber tip is monitored in real time using a CCD camera combined with a 100× objective len (Nikon, NA = 0.9) that collects transmitted light. Moreover, by adjusting the illumination wavelength or introducing a controlled lateral offset during coupling light into the fiber input facet, the guided mode composition inside the MMF can be rapidly reconfigured. This allows for controllable axial displacement of the optical potential, enabling dynamic axial manipulation of samples.

## 2.2 Micro-lens Structure Design and Fabrication

As illustrated in Fig. 1(c), the designed micro-lens structure features a truncated cone-shaped body topped with a hemispherical concave cavity on the top, supported by a reinforced base that enhances mechanical stability and adhesion to the fiber facet. The radius of the bottom is 52.5 μm that covers the 105 μm-diameter core of the MMF. It is noted that the sloping sidewall of the cone-shaped body can reflect and redirect high-order modes propagating along the outer region of the core, thus increasing the effective convergence angle and contributing to the formation of a high-NA focal spot. Additionally, the top concave cavity can diverge the low-order modes propagating along the fiber center, reducing the scattering force contributed by these modes and enhancing trapping stability. Overall, the micro-lens effectively reshapes and tightly focuses the ring-shaped high-order guided modes emitted from the MMF, enabling stable trapping at the focal region.

In experiment, the micro-lens structures were fabricated using two-photon polymerization (TPP)–based three-dimensional nanofabrication with spatial resolution far beyond that achievable by conventional photolithography [21–23]. All structures were initially designed in SolidWorks and then fabricated using a Nanoscribe GT system. The platform operates with a 780 nm femtosecond pulsed laser (pulse width 120 fs, repetition rate 80 MHz) and supports high-precision (<60 nm) direct laser writing on the fiber facet. A representative cross-sectional SEM image of the fabricated micro-lens is shown in Fig. 1(d), with structural dimensions as follows: the radius of the hemispherical cavity Rc =15μm, the base radius Rb=52.5 μm, the lower structural radius rb=40 μm, the central height h=68.97 μm, and the sidewall angle of 15° relative to the vertical axis.

# 3. Theoretical and Experimental Results

## 3.1 Optical focal Field and Forces

To evaluate the focusing and trapping performance of the fiber-end micro-lens structure at different wavelengths, three-dimensional finite-difference time-domain (3D FDTD) simulations were conducted using Lumerical FDTD Solutions (2020 R2.4). The total simulation region is 40×40×40 $\mu m^3$, with a minimum mesh size of 20 nm. To be consistent with the experimental conditions, two light sources (a continuous-wave laser at 532 nm, and a pulsed laser with a central wavelength of 800 nm and pulse width of 100 fs) were used in FDTD simulations. To generate unpolarized illumination to the micro-lens, two separate simulations were performed: one with multiple random-phase sources superimposed under x polarization, and the other under y polarization. This approach approximates an unpolarized light source under mixed mode conditions. The resulting electromagnetic field distributions from the two simulations were then averaged. Owing to the rotational symmetry of the micro lens, this method effectively reproduces the polarization independent optical response. The total incident power was set to 35 mW to match the experimental conditions. The refractive index of the micro-lens was set to n = 1.54 at 532 nm and n =1.53 at 800 nm, corresponding to the IP-DIP photoresist used in experiment. The trapped objects were modeled as spherical particles with a diameter of 4 μm and a refractive index of 1.40, similar to a yeast cell immersed in water [24]. Optical forces were calculated using the Maxwell stress tensor method [25], and the corresponding optical trapping potential was obtained according to the formulation: $U(r_0) = -\int_{\infty}^{r_0} F(r) \cdot dr$[26], which provides an accurate characterization of the trap depth and confinement strength. The FDTD simulation results at both wavelengths of 532 nm and 800 nm are presented in Fig. 2, illustrating the focused field distribution generated by the fiber-end micro-lens, along with the axial and transverse optical forces and the corresponding trapping potentials acting on particles near the focal region.

The field intensity distributions shown in Figs. 2(a) and 2(b) demonstrate that the designed micro-lens structure produces elongated focal spots along the axial direction for both 532 nm and 800 nm wavelengths. To evaluate the focusing capability of the micro-lens, the effective NA of the focusing beam was calculated based on the radial spatial frequency distribution ($k_\perp/k_0$) [27], where the radial wavevector component is defined as:

$$k_\perp = \sqrt{k_x^2 + k_y^2}\,, \quad k_0 = \frac{2\pi}{\lambda} \tag{1}$$

Here, $k0$ is the wavenumber of the incident light, and kx and ky are the in-plane components of the wavevector. Each transverse wavevector corresponds to a specific propagation angle θ, given by:

$$\sin\theta = \frac{k_\perp}{k_0} \tag{2}$$

By performing a 2D Fourier transform of the total field distribution E(x, y) at two axial cross-sections, one located 5 nm above the micro-lens tip and the other 5 nm above the focal position, the normalized radial energy spectrum ($k_\perp/k_0$) can be obtained according to Eqs. (1-2). The radius corresponding to 95% of the cumulative energy is taken as the effective cutoff of the radial spectrum [28], which defines the maximum achievable propagation angle θmax of the beam. The effective NA is then calculated as

$$NA_{eff} = n\sin\theta_{max} = n\left(\frac{k_{\perp,eff}}{k_0}\right) \tag{3}$$

Here, n is the refractive index of the surrounding aqueous medium. The corresponding radial energy spectra, shown in the insets of Figs. 2(a) and 2(b), exhibit pronounced high-frequency components ((see Supplementary Discussion 1 for details)). The calculated effective NA in water reaches approximately 0.71 under 800 nm wavelength and 0.78 at 532 nm wavelength. Such high NA values are remarkable for a MMF, where the typically low output coherence and the random phases of guided modes generally hinder the formation of tightly focused spot.

Furthermore, the calculated optical forces on particles near the focal region are shown in Figs. 2(c) and 2(d), depicting the axial force ($F_z$) and the transverse force ($F_x$), respectively. It is observed that the $F_z$ at 532 nm exhibits a significantly higher maximum compared to 800 nm, and the $F_x$ gradient at 532 nm is steeper, indicating a stronger confinement. The zero crossings of the force curves in Fig. 2(c) correspond to the equilibrium trapping positions, located approximately 15 μm from the micro-lens tip for 800 nm and 17 μm for 532 nm, with the shorter-wavelength trap slightly farther from the tip. These results demonstrate that the high-NA micro-lens can generate sufficient optical gradient forces to drive particles toward the focus and confine them at the axial equilibrium position.

The optical potential distributions at the trapping plane are shown in Figs. 2(e)–(f). For both wavelengths, deep trapping potentials can be observed, guiding particles radially toward the potential minimum. It is well known that when the depth of trapping potentials exceeds ~10 kBT (kB is Boltzmann constant, T is absolute temperature), Brownian motion of particles can be effectively overcome, enabling stable optical trapping [26, 29]. Quantitatively, the depth of trapping potential reaches 134.748 kBT at 532 nm and 132.372 kBT at 800 nm, both sufficient for long-term trapping of micron-sized targets (e.g., yeast cells) over a large spatial range. These results clearly demonstrate that the designed micro-lens allows broadband optical focusing and trapping for MMF-based tweezer systems, and has potential for wavelength-dependent control of trapping position and potential depth.

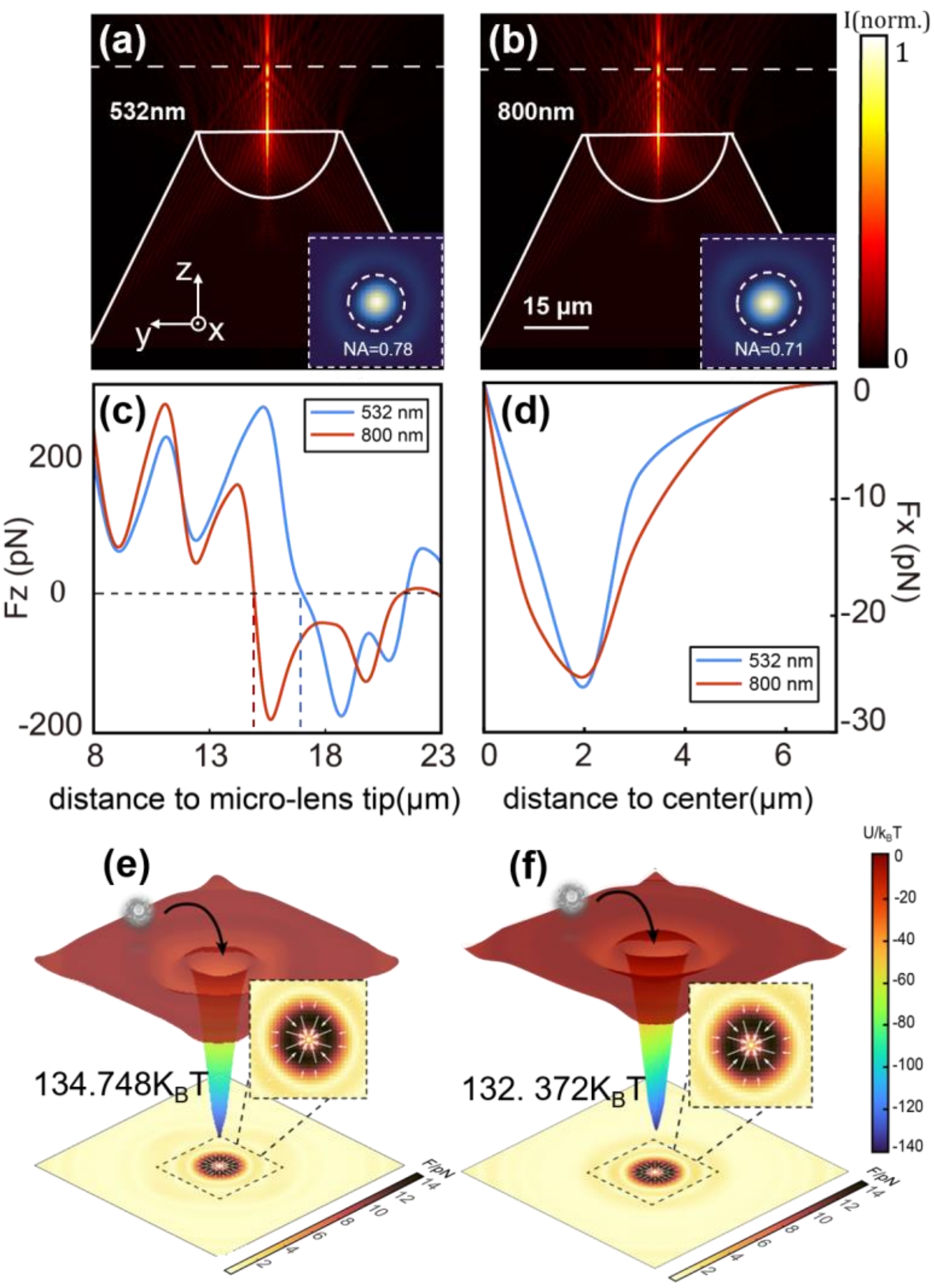

Fig. 2.: FDTD simulation results of the micro-lens structure under different wavelengths. (a, b) Electric field distributions at the top region of micro-lens for the wavelengths of 532 nm and 800 nm. The bottom-right insets show the focal spot distributions obtained via inverse FFT fitting, with the effective NA=0.78 for 532 nm and NA=0.71 for 800nm. (c, d) Axial and transverse optical forces on particles at different positions. Red curves correspond to 800 nm, blue curves to 532 nm, and the black dashed line indicates the equilibrium position. The red and blue dashed lines denote the trapping positions for each wavelength. (e, f) Optical potential distributions under 532 nm and 800 nm excitation, along with schematic diagrams of force distribution. The upper panels show the potential distributions in the XY plane, and the lower panels depict the in-plane force distributions at the trapping plane for each wavelength, with white arrows indicating the directions of local forces.

### 3.2 Experimental trapping and dynamic control of yeast cells

Using the high-NA micro-lens structures fabricated on the MMF facet, we experimentally investigated the stable trapping and dynamic manipulation of yeast cells. As a representative microbial sample, yeast cells (Pichia pastoris, cell diameter of 4~6μm) are widely used in biological research, drug screening, and single-cell analysis [30, 31], making them an ideal target for validating the high-precision MMF optical tweezer platform.

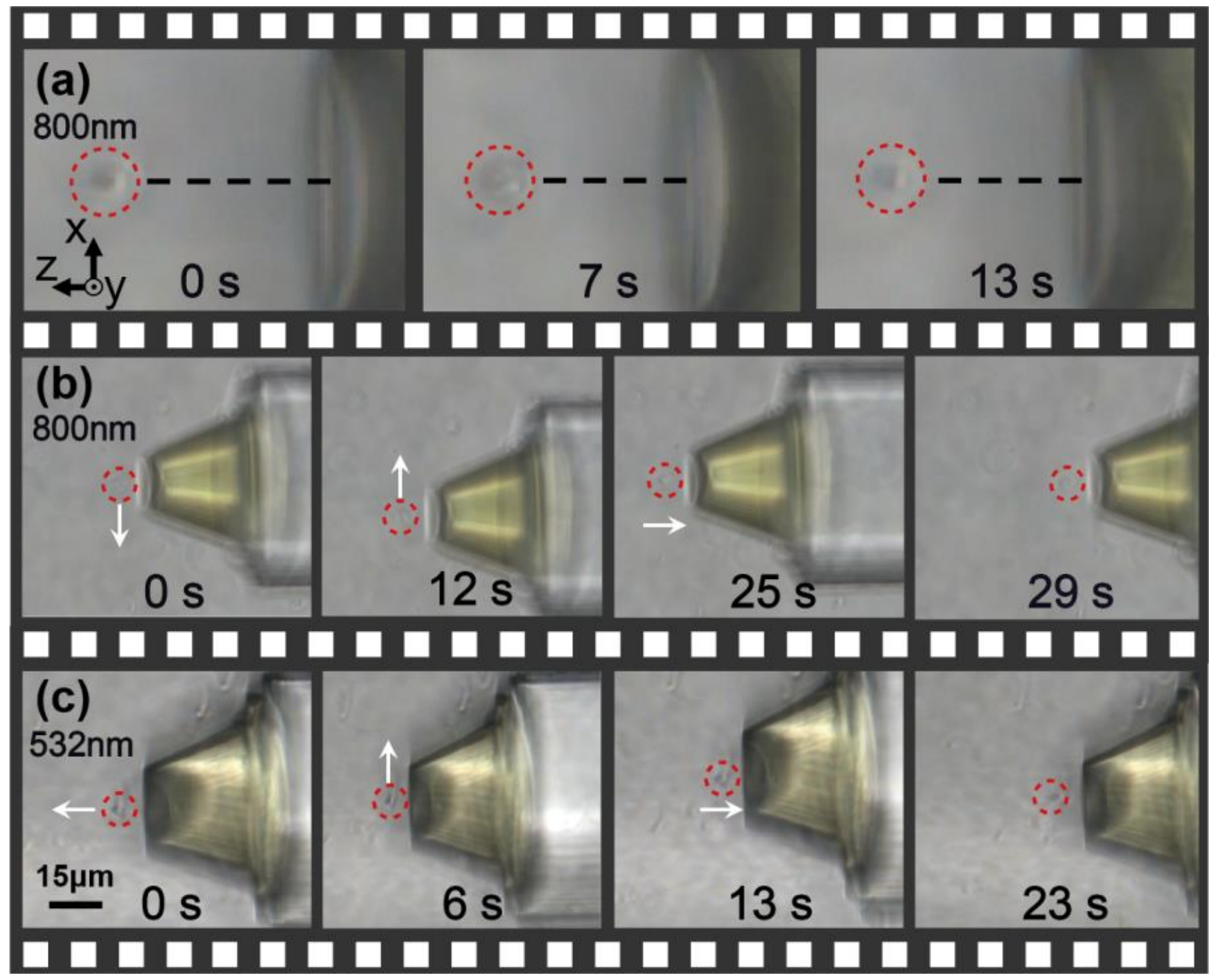

Fig. 3： Trapping of yeast cells using the integrated micro-lens structures on MMF facet. (a) Time-lapse images showing the trapping process of a yeast cell approaching the micro-lens under 800 nm wavelength light source. (b, c) Trapped cell trajectories following the MMF movements to evaluate trap stability under 800 nm and 532 nm wavelength sources, respectively. Red dashed circles mark the trapped single cell, and the white arrows indicate the movement direction. Scale bar: 15 μm.

Time-lapse images in Figs. 3(a–c) illustrate the trapping performance of the integrated micro-lenses on yeast cells using both 800 nm and 532 nm illumination (also shown in Supplementary Video S1-S3). In Fig. 3(a), as the fiber approaches the cell, freely diffusing yeast cell gradually moves toward the fiber tip under optical gradient forces. At the initial trapping stage (t = 0 s), the cell exhibits slight axial defocusing, and then it is drawn into the focal region and finally achieves stable trapping near the focus at (t = 13 s). Under 800 nm illumination, stable trapping can be achieved at a laser power range of 40~120 mW. As shown in In Fig. 3(b), the cell (in red dashed circle) remains stably confined over 29 s, even when the fiber is intentionally moved along x-direction at speed of 10 μm/s during (t = 0~12 s) and along z-direction at 16 μm/s during (t = 25~29 s), demonstrating the stability from the deep trapping potential. Under 532 nm illumination (Fig. 3(c)), cell trapping can be achieved at a lower power range (20~60 mW) but exhibits stronger spatial confinement. As shown in Fig. 3(c), during the fiber quickly moves along z-direction at 25 μm/s (t = 0~6 s) and along x-direction at 32 μm/s (t = 13~23 s), the trapped single cell still maintains at the focal region, owing to the higher effective NA (Fig. 2(a)) and deeper potential well (Fig. 2(e)) at the 532 nm wavelength. These results verify that the micro-lens enables stable, reproducible trapping across both visible and near-infrared femtosecond light sources, highlighting the inherent advantages of MMFs for multi-wavelength operations. Compared to conventional SMF-based optical tweezers, which are typically narrowband, this MMF approach offers superior wavelength adaptability and trapping flexibility.

In conventional SMF tweezers, the output field is dominated by the fundamental guided mode of the SMF, which generally precludes dynamic modulation via optical modes. In contrast, MMFs support multiple guided modes, providing additional degrees of freedom to tailor the spatial distribution of the output optical field. Figure 4 further illustrates the physical mechanism, numerical analysis, and experimental validation of axial cell manipulation achieved through on-axis and off-axis coupling between the incident light and the MMF. As schematically shown in Fig. 4(a), adjusting the input coupling conditions excites distinct guided modes within the MMF, leading to either a ring-shaped intensity profile associated with high-order modes (off-axis coupling condition) or a Gaussian-like optical field dominated by low-order modes (on-axis coupling condition). The corresponding experimental measured output light intensity pattens of the two cases through the MMF without micro-lens are presented in the insets of Fig. 4(a). After reflection by the integrated micro-lens, these high- and low-order modes are focused at different axial positions beyond the fiber tip, enabling controllable displacement of the trapping potential along the propagation direction. As illustrated by the schematic on the right side of Fig. 4(a), under on-axis coupling, the optical field has a smaller transverse radius and is dominated by low-order modes. Consequently, the rays are reflected at positions farther along the sidewall of the microlens, leading to a focal spot located farther away from the fiber tip. In contrast, under off-axis coupling, the excitation of high-order modes results in a larger ring-shaped field distribution, causing the rays to be reflected closer to the microlens base, and thus producing a focal spot located closer to the fiber end face.

The corresponding FDTD simulated results are shown in Fig. 4(b). When the beam is laterally offset (off-axis coupling condition), the FDTD results indicate that the peak of the focal field generated by the micro-lens is close to the fiber facet. In contrast, when the beam is aligned with the fiber axis (on-axis coupling condition), the peak position of the focal field becomes farther away, just as the prediction in Fig. 4(a). In the experimental verification (Fig. 4(c)), a single yeast cell was firmly trapped and then moved forward by $\Delta Z \approx 6.5$ μm during the shift from off-axis to on-axis coupling conditions, consistent with the FDTD simulation, confirming that the modulation of input coupling conditions provides an effective dynamic control method for the MMF optical tweezer. Detailed experimental process is shown in Supplementary Video S4. Compared to SMF tweezers, which are usually limited to a single guided mode and a fixed focal position, the MMFs provide enhanced spatial control through mode diversity with high flexibility, offering a new approach for multi-position and switchable optical trapping.

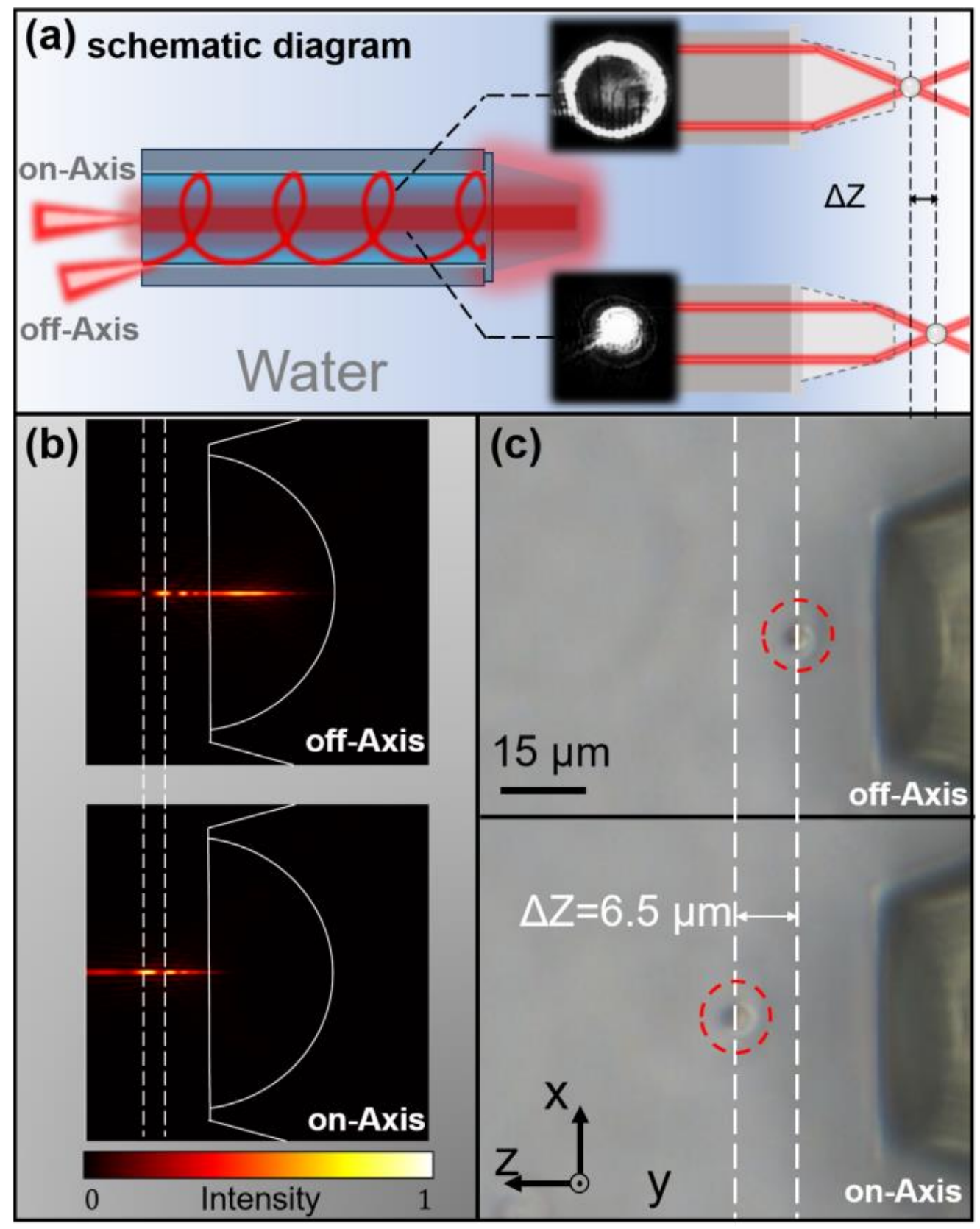

Fig. 4： Axial dynamic manipulation of cells via coupling between the incident light and the MMF. (a) Schematic of the modulation principle(see Supplementary Discussion 2 for details). By switching between on-axis and off-axis coupling at the input facet of the MMF, the transmission mode within the MMF is reconfigured, generating ring-shaped intensity profiles with varying radii at the fiber tip. After focusing by the integrated micro-lens, distinct transmission modes produce axially shifted focal positions, enabling dynamic axial control of the trapping location (ΔZ) of yeast cells. The right panel insets illustrate the experimentally measured output light patterns formed under on-axis and off-axis coupling conditions. (b) FDTD simulation results of the shifted focal positions at the 800 nm illumination under off-axis (top) and on-axis (bottom) conditions. (c) Experimental demonstration of axial dynamic manipulation of a single yeast cell at the 800 nm illumination. The red dashed circle marks the trapped cell, and while dashed lines indicate the different focal positions. An axial trapping displacement of (ΔZ = 6.5 μm) is achieved. Scale bar: 15 μm.

## 4. conclusion

In summary, this work overcomes a long-standing limitation of MMFs, namely, the modal scrambling that typically impedes the formation of high-intensity focal fields. By integrating a high-NA micro-lens structure onto the fiber tip, we demonstrate a fully MMF-based optical tweezers system capable of generating deep optical potential wells and achieving robust three-dimensional trapping under multi-wavelength conditions. Using 532 nm continuous-wave and 800 nm femtosecond lasers as representative light sources, both produce tightly confined focus with strong gradient forces. Although the overall potential depths are comparable, the 532 nm excitation yields a deeper and narrower potential well, corresponding to better performance in experiment, whereas the 800 nm operation reduces optical absorption and photodamage in biological samples, rendering it more suitable making long-term manipulation [32]. Furthermore, by introducing controllable light coupling at the fiber input, selective excitation of guided modes is achieved, resulting in distinct axial focal positions beyond the fiber tip. This mechanism enables all-optical modulation of the trapping position and optical force landscape through the selective guided modes within MMFS.

Compared to conventional SMF tweezers, the proposed fully MMF tweezers platform offers distinct advantages in focusing capability, operational wavelength range, and dynamic regulation capability. The involvement of designed micro-lens significantly enhances the effective numerical aperture (NA > 0.7) for both wavelengths (532 nm and 800 nm), thereby forming deep trapping potential wells. Meanwhile, the broadband transmission of MMFs supports stable trapping and manipulation across different wavelengths various a single platform. The high sensitivity of the output focal field to input coupling conditions further provides tunability over trap positions, reducing the requirement for additional beam-shaping devices.

The incorporation of MMFs renders the proposed optical tweezer system highly suitable for integrated "trap-and-spectroscopy" applications. For instance, it enables simultaneous trapping and targeted drug delivery [33,34], Raman spectroscopic analysis [35], and even time-resolved fluorescence lifetime measurements when combined with ultrafast excitations [36]. While such multifunctional capabilities generally necessitate intricate optical routing or the use of multiple fibers in conventional optical fiber tweezer systems, the proposed design can achieve these functions through a single MMF configuration, hereby significantly simplifying the overall system architecture. Moreover, the dynamic manipulation enabled by on/off-axis coupling can contribute to applications such as probing cellular mechanical responses in viscoelastic media, mapping local microenvironment viscosity, and extracting dynamic cellular parameters [37,38]. Collectively, these findings highlight the potential of fully

MMF-based optical tweezers to realize versatile functions, offering a compact and powerful platform for fiber-integrated biological manipulation and multidimensional optical sensing.

## Supporting Information

- Determination of the Effective NA from the Angular Spectrum, Beam Profile Evolution under Off-axis Coupling Modulation (PDF)
- Video 1: Stable Trapping Process of a Yeast Cell under 800 nm Excitation. (MP4).
- Video 2: Trapping of a Yeast Cell Using an 800 nm Femtosecond Laser. (MP4).
- Video 3: Trapping of a Yeast Cell under 532 nm Excitation .(MP4).
- Video 4:Experimental Demonstration of Axial Dynamic Manipulation of a Single Yeast Cell under 800 nm Excitation. (MP4).

## Acknowledgement

The authors acknowledge the Photonics Center of Shenzhen University for technical support.

**Conflicts of interest:**There are no conflicts to declare.

**Funding:** This work was partially supported by National Natural Science Foundation of China (62575184, 62375177, and 12534017); Shenzhen Science and Technology Program (RCJC20210609103232046); Shenzhen Medical Research Fund (D250403001); Project of DEGP(2024ZDZX2019). Research Team Cultivation Program of Shenzhen University (2023QNT014); Shenzhen University 2035 Initiative (2023B004).

**Author contributions:** Y.S. conceived the outlined and accomplish the manuscript. C.Z.and Y.Z. provided the experimental ideas. J.P.,S.C. and H.X. provided simulation guidence. Y.C. and Q.J. assisted with data curation. C.J. and Y. S. (co-corresponding authors) reviewed and edited the manuscript.

**Data Availability:**Data underlying the results presented in this paper are not publicly available at this time but may be obtained from the authors upon reasonable request.